\documentclass[12pt,letterpaper]{JHEP3}
\usepackage{epsfig}

\def\IZ{{\mathbb Z}}
\def\IR{{\mathbb R}}

\def\I1{{\mathbb 1}}
\def\lsim{\mathrel{\rlap{\lower 4pt \hbox{\hskip 1pt $\sim$}}\raise 1pt \hbox
    {$<$}}}
\def\gsim{\mathrel{\rlap{\lower 4pt \hbox{\hskip 1pt $\sim$}}\raise 1pt \hbox
    {$>$}}}

\def \da9{D9-\bar{D}9}

\def \n{\noindent}
\def \dim{{\rm dim}}

\def \dim{{\rm dim}}
\def \beq{\begin{equation}}
\def \eeq{\end{equation}}
\def \da9{D9-\bar{D}9}

\title{D-brane Anti-brane Annihilation in an Expanding Universe}
\author{Mahbub Majumdar$^1$ \& Anne-Christine Davis$^2$ \\

$^1$ Theoretical Physics Group\\ Imperial College, Blackett Lab\\ Prince
Consort Road, London SW7 2BW, UK\\
$^2$ Department of Applied Mathematics and Theoretical Physics, \\
   Centre for Mathematical Sciences\\ University of Cambridge,
   Wilberforce Road, Cambridge, CB3 0WA, U.K.\\
Emails: {\tt m.majumdar@ic.ac.uk, A.C.Davis@damtp.cam.ac.uk}

}

\abstract{The time-varying density of D-branes and anti-D-branes
in an expanding universe is calculated.  The D-brane anti-brane
annihilation rate is shown to be too small to compete with the
expansion rate of a FRW type universe and the branes over-close
the universe. This brane problem is analogous to the old monopole
problem.  Interestingly however, it is shown that small dimension
D-branes annihilate more slowly than high dimension branes. Hence,
an initially brany universe may be filled with only low dimension
branes at late times.  When combined with an appropriate late
inflationary theory this leads to an attractive dynamical way to
create a realistic braneworld scenario.

 }

\date{\today}

\begin{document}


\section{\bf Introduction}

In a previous article we showed how $D$-branes may have been
produced in a tachyonic phase transition at an energy near the
string scale~\cite{paper2}. While providing a well-defined non-perturbative
method of generating branes, this mechanism suffers from the
typical problem common to all topological defect theories -- too
many defects are produced.

The mass of a $D$-brane is proportional to its volume and is
inversely proportional to the closed string coupling $g_s$.  Thus
non-compact branes at weak coupling are extremely heavy.  Compact
branes which wrap compact cycles of a spacetime $M$ are less
heavy, but are still much heavier than fundamental string states
like electrons and heavy gauge bosons, etc.  Thus, a gas of
branes, whether compact or non-compact, is  likely to cause the
energy density of the universe to exceed the critical energy
density and turn an expanding universe into a contracting
one~\cite{bv2}.

However, if $g_s \rightarrow 0$, branes do not pose many
cosmological problems because they then decouple from gravity in
bulk. This is albeit that the mass $m_{Dp} \sim V_{p}/g_s$
diverges.  Gravity couples to D-branes through closed string
emission and absorption.  The ``vertices'' (in field theory
parlance) for such interactions are generated by closed string
vertex operators, $V({\bf k})$, which are proportional to $g_s$.
Thus as $g_s \rightarrow 0$, the vertex operators ``tend to''
zero, closed string emission/absorption is suppressed, and gravity
decouples from the branes.  This is similar to the fact that the
vertex for electron-photon interactions is proportional to the
fine-structure constant $\alpha$; and if $\alpha \rightarrow 0$
then electrons and photons no longer interact.  Thus
interestingly, heavy branes do not affect the gravitational
dynamics of the universe in this context.

This can be seen in a different manner. The perturbation $\delta
g_{00}$ of a massive object on the surrounding spacetime in linearized
gravity is given by the Newtonian potential,

\begin{equation}
\delta g_{00} = - \frac{G_{d} m_{Dp}}{r^{d-2}} \longrightarrow 0~~~
{\rm if}~g_s \rightarrow 0
\label{g00}
\end{equation}

\n because the $d+1$-dimensional Newton's constant varies as
$G_{d+1} \sim g_s^2 $ and $m_{Dp} \sim 1/g_s$.

However, for reasonable values of $g_s$ (which is related to the
Yang-Mills coupling on the branes) branes will distort any
cosmological spacetime.  It is therefore imperative to dilute their densities
via brane anti-brane annihilation, inflation, or something similar.
In this paper we try to determine how brane anti-brane annihilation
affects brane densities.

If extra compact directions exist then the density of branes and
anti-branes at late times will significantly depend on the process of
decompatification of the large dimensions or compactification of the
small dimensions, the initial topology of spacetime, and more
generally the evolution of the compact directions. Our results can be
summarized as follows.

Suppose that all of the compact directions are stabilised before
brane production and that four non-compact directions are
expanding. Then because of strong brane anti-brane interactions,
branes filling the non-compact directions and separated only in
the compact directions will quickly annihilate.  Such branes will
not feel the cosmic expansion which typically causes interactions
to freeze out below some temperature.  Thus branes with $p\ge3$
which completely fill the non-compact directions and possibly wrap
some of the compact directions will not pose any cosmological
problems. Interactions between branes with $p<3$ will be frozen
out, and these branes may pose severe cosmological problems.
Branes with $p \ge 3$ which do not fill the non-compact directions
will pose problems as well.

If all of the dimensions, compact and non-compact, are expanding
at the time of brane production then branes of all
dimensionalities will soon dominate the energy density of the
universe and cause it to collapse.  This is because brane
interactions will quickly be frozen out and they will be unable to
annihilate. If none of the dimensions have been decompactified,
such that all the directions are compact but expanding as in the
Brandenberger-Vafa model, then a variety of possibilites are
possible. If the wrapped branes can generate a positive pressure
and confine the compact directions via some sort of  dilaton
dynamics, then all of the wrapped branes may  annihilate. This
will ameliorate any brane problem but also prevent wrapped branes
from allowing  four compact directions to grow large.  If the
compact directions are not stabilized, then interactions will be
frozen out and branes will be littered throughout spacetime.

One interesting, but expected result from our calculations is that
high dimensional branes annihilate more efficiently and drop out
of equilibrium later. This is essentially because they interact
more vigorously with other branes  and matter in the bulk. This
means that a Type IIB universe initally filled with branes may
migrate to a universe filled with only a few $D3$ branes and many
$D1$ strings. This may be very attractive from a braneworld point
of view.

The plan of this paper is as follows. First we consider the idealized
case of non-interacting branes in a static universe; this is
essentially a reworking of the heuristic arguments of \cite{bv1} from
a somewhat more algebraic point of view. Then we consider the case of
interacting branes in a static universe.  Next, we move onto the more
realistic case of interacting branes in an expanding
universe. Finally, we end with a catalog of how our results apply to
scenarios with differing numbers of compact directions, expanding
directions, wrapped directions, etc..

\subsection{Non-interacting branes in Flat Space}\label{flatspace}

If we ignore long-range interactions and the cosmic expansion rate
$H^{-1}$, then the annihilation rate of a system of branes in a static
spacetime $M$, should depend only on the distinguishing features of
the branes such as their dimensionality $(p+1)$, and the geometry,
topology and dimensionality of the spacetime $M$.

Consider a non-coincident $Dp$ brane and $\bar{D}p'$ anti-brane.
Label them as brane $A$ and brane $B$ and their worldvolumes as
$A$ and $B$. $A$ and $B$ will annihilate if they are coincident,
parallel, and if $p=p'$.  Anihilation occurs because tachyons
arise  which eventually condense to the minima of their potential.
The minima for coincident anti-parallel branes corresponds to the
disappearance/confinement of all open string degrees of freedom --
i.e. annihilation~\cite{sen}.  Tachyons also arise if two branes
intersect at angles~\cite{angles}.  However, the minima of the
tachyon potential in that case correspond to changes in the
geometry of the branes and not annihilation.

A $Dp$ brane and $\bar{D}p'$ anti-brane will only meet each other if
their spacetime paths intersect at some point in time -- i.e. if their
worldvolumes intersect. We have the following set theory identities

\begin{equation}
A + B  =  A\setminus B + B\setminus A + A \cap B
\label{settheory}
\end{equation}

\n
where

\begin{eqnarray}
A\setminus B & = & A - A \cap B \nonumber\\
B\setminus A & = & B - A \cap B.
\end{eqnarray}

\n
This allows us to write

\begin{eqnarray}
\dim(A+B) & = & \dim(A -A \cap B) + \dim(B - A\cap B) + \dim(A\cap
B)\nonumber\\ & = & \dim(A) + \dim(B) - \dim(A\cap B).
\label{dimproblem}
\end{eqnarray}

\n where $\dim(A\cap B)$ is appropriately defined to ensure that the
$\dim$ operator is linear.\footnote{For example, if $ \dim(A) = \dim(B)
> \dim(A+B)/2$ and the worldvolumes $A$ and $B$ are non-intersecting
and parallel, then (\ref{dimproblem}) would imply that $ \dim(A\cap B)
>0$ which cannot be true.  This is ameliorated by defining the
intersection of the worldvolumes of two parallel objects to be
intersecting in $\IR^n$.  This is reasonable because generic such
worldvolumes will not be parallel and will intersect. Alternatively,
one can work in projective space where for example, two parallel
lines always intersect.} We now use the definitions

\begin{eqnarray}
\dim(A + B)  \le  D;~~
& \dim(A)  =  p+1; &
~~\dim(B)  =  p'+1
\end{eqnarray}

\n
where $D$ is the dimensionality of the spacetime.  We then conclude that

\begin{equation}
\dim(A\cap B) \ge (p+1) +(p'+1) -D
\label{intersection}
\end{equation}

Thus, brane $A$ and anti-brane $B$ will surely meet at some point in
time if $(p+p')> D-2$.  Equation \ref{intersection} provides a {\em
lower bound} on the dimensionality of the intersection, and the actual
dimensionality of the intersection will usually be larger.  For
example, if $(p+p') = D-2$, then $\dim(A\cap B) \ge 0$.  The only way
$\dim(A\cap B) = 0$ and $A \cap B = \emptyset $ can occur is if the
worldlines of $A$ and $B$ are roughly parallel - an unlikely
situation.  Otherwise, they will intersect in a non-zero dimensional
space.  In the special case when $p=p'$, and the $Dp$ and $\bar{D}p$
are parallel, any intersection of $A$ and $B$ will be $p+1$
dimensional. This is crucial, because annihilation requires the brane
and anti-brane to be coincident and that $\dim(A \cap B) = p+1$.

In string theory, $D=10$.  Thus, $D3$ branes will (almost
surely)\footnote{In the measure theoretic sense.} not meet. But, $D5$
and $D7$ branes will meet and annihilate. $D3$ branes and lower dimensional branes
may meet. But the number of such configurations is a set of measure
zero.  For example, intuition tells us that two non-interacting low
dimension branes can on rare occasions simply ``run into each other.''
This will occur if brane $A$ and brane $B$ live in a smaller space
$W\subset M$, which is a proper subset of $M$.  Then effectively, $D =
\dim(W)$. For example, if the worldlines of two $D0$ branes lie in a 2D
plane in a 10D dimensional space, then $W = \IR^2$; $D=2$; and unless
the worldlines are parallel, the branes will meet. The probability of
two randomly moving objects lying in a proper subset of $\IR^{10}$ is
small. This would require that a subset of the transverse coordinates
of  $A$ in $M$ and $B$ in $M$ coincide exactly.  The number of such
coordinates is $\dim(M-W)$. Thus, the probability, $P$ of lying in
the subset is of order

\begin{equation}
P \sim \left(\frac{1}{{\rm vol}(\IR)}\right )^{\dim(M-W)}
\end{equation}

\n which is vanishingly small.  The actual probability will be larger,
since $A$ and $B$ cannot be completely random surfaces.  They must lie
within their respective lightcones.

However, if the transverse dimensions to $W$ are compact with a compactification radius of $l$, then

\begin{equation}
P \sim \left ( \frac{1}{l}\right )^{\dim(M-W)}
\end{equation}

\n which may still be small, but does not vanish. Thus,
compactification helps branes meet.  But eq. \ref{intersection} is
still the principal criterion determining when non-interacting $Dp$
and $Dp'$ branes will meet.

However, in reality branes are rarely non-interacting.  Only parallel
branes of the same dimension with the same orientation feel no
attraction or repulsion to each other.  An inter-brane potential
exists for branes at angles, branes of different dimensions, and
branes and anti-branes~\cite{joebigbook,angles,narain}.

\subsection{Interacting Branes in Flat Space}

Recall that the charge of a brane relative to another brane depends on the
relative orientation.  The extreme case is when a brane is rotated
by an angle $\pi$ relative to another brane.  This corresponds to an
anti-parallel or anti-brane.  Just as particles and anti-particles
feel a mutual attractive force, so do branes and anti-branes.  Two
branes with zero relative rotation angle are parallel branes.  Thus,
it is not surprising that the intermediate case of a rotation by $0<
\theta < \pi$ interpolates between no force and the large attractive force
of brane anti-brane interaction; i.e. branes at angles feel forces.
However, at a few special angles the force vanishes and the
configuration is stable.

Branes of different dimension also feel forces because open string
excitations connecting the branes are tachyonic.  For example, a
$Dp$ is attracted to a $D(p+2)$.  The force disappears once the
$Dp$ brane lies on the worldvolume of the $D(p+2)$ brane. A $Dp$
is repulsed by a $D(p+6)$ brane, and unlike the $Dp-D(p+2)$ system
does not flow toward a stable system~\cite{narain}. $Dp-D(p+4)$
brane configurations are supersymmetric and feel no forces
~\cite{joebigbook,angles}.

We will focus on the force between parallel branes and anti-branes of
the same dimension. Only such systems can decrease the
brane/anti-brane number densities in the universe.

The potential $\Phi(r)$ between a  $Dp$ brane at a distance $r$ away from a  $\bar{D}p$ is the analog of the gravitational
potential, $\phi(r)$ in four dimensions

\begin{equation}
\phi(r) = - \frac{G_4 m}{r}
\;\;\;\;\;\; \rightarrow
\;\;\;\;\;\; \Phi(r) = {\rm constant} \frac{G_{10} \tau_p}{r^{7-p}}
\end{equation}

\n where $G_{D}$ is Newton's constant in $D$ spacetime dimensions. In
place of the mass, the potential contains the tension $\tau_p$. The
quantity $G_D \tau_p$ is analogous to the effective gravity of cosmic
strings, $G \mu$, where $\mu$ is the string tension. It measures the
gravitational strength of the branes. The potential energy $V(r)$, of
a brane and anti-brane is therefore $V(r) = m_{Dp} \Phi(r)$ in analogy
with the purely gravitational case. Since $\kappa^2 = 8 \pi G_{10}$,
we can write $V(r)$ as~\cite{susskind}

\begin{equation}
V(r) = - \beta \frac{\kappa^2 \tau_p m_{Dp}}{r^{7-p}} \equiv -\frac{h^2}{r^{7-p}}
\label{PE}
\end{equation}

\n
Here, $\beta$ is a numerical factor characteristic of
D-branes~\cite{susskind,paper3}.

\begin{equation}
\beta = \pi^{-(9-p)/2} \Gamma(\frac{7-p}{2})
\end{equation}

A brane and anti-brane in relative motion will have velocity and
acceleration dependent potentials as well~\cite{lifschytz,periwal}.
For example, the largest velocity dependent piece is

\begin{equation}
\Delta_v V(r) = - \frac{v^2}{2} \frac{h^2}{r^{7-p}} +
O(\frac{v^4}{r^{7-p}})
\end{equation}

The first piece appears because of supersymmetry breaking and the
second is present in susy and non-susy cases.  Since both terms
change the potential energy in (\ref{PE}) by no more than a small
numerical factor (of the same sign), we will ignore them. We will
also ignore acceleration terms.  Note, we have left out the rest
mass $2m_{Dp}$ in (\ref{PE}).

If some directions are compact then the brane inter-brane potential
will depend on the average distance between the branes and
anti-branes.  If the distance is comparable to the size of the compact
directions then the branes will notice the compactness of the
extra-dimensional space.  Hence, to calculate the effective brane
anti-brane potential, the effects of an infinity of image charges will
have to be added.  This will soften the potential.  However, the
inter-brane separation is unlikely to be close to the compactification
scale $r_c$, as a gas of branes with a non-neglible density $n$, will have an
inter-brane seperation of $n^{-(d-p)}$ which will usually be much
smaller than $r_c$ -- and this is the case that we will
investigate.

Because of long-range interactions the naive dimension-based arguments
in the previous section do not hold for more general brane systems.
Long rang forces lead to infinite scattering cross-sections.  Thus
branes and anti-branes will disappear due to annihilation if the
universe remains static for a sufficiently long time and if
annihilation doesn't require impossible topology change of any wrapped
D-branes (impossible winding number transitions).

\subsection{Branes in an Expanding Universe}

We can calculate the brane and anti-brane abundances in an
expanding universe as in~\cite{preskill,zelda}.  First assume that
the branes and anti-branes form dilute gases.  This assumption
allows us to ignore correlations between brane and anti-brane
positions arising from the physics of a tachyonic phase transition
which may have created them.  Next, assume that the brane density
equals the anti-brane density to ensure charge neutrality. The
time evolution of the branes/anti-branes is then given by

\begin{equation}
\frac{dn}{dt} = -  nH(d_H-p_H)  -  \Gamma n^2.
\label{dilute}
\end{equation}
\n Here $\Gamma$ characterizes the annihilation process; $dn/dt$
is the time derivative of the density of branes $n$; and $d_H$ is
the number of expanding dimensions. The constant $p_H$ is the
number of expanding directions which the branes fill. If $p_s$ is
the number of static dimensions which the branes fill (non-compact
or compact), then

\begin{equation}
p = p_H + p_s
\end{equation}

\n The quantity $(d_H-p_H)$ represents the number of expanding
dimensions in which the branes move.  The first term in (\ref{dilute})
represents the dilution effect of cosmic expansion.  The second
represents brane anti-brane annihilation.

The $d_H$ expanding dimensions may be compact or non-compact.
Suppose that they are all non-compact and that the remaining
$d-d_H$ spatial dimensions are compact.  Then branes living only
in the compact directions will still be diluted because their
inter-brane separations in the non-compact directions will grow.
The only way branes in the compact directions will not feel the
first term in (\ref{dilute}) is if they are all coincident in the
non-compact directions which is very unlikely, or they competely
fill the extra-dimensions such that $d_H = p_H$.  For example, if
$d_H=3$, then a gas of $D3$ branes all lying in the four
non-compact directions and separated only in the compact
directions will not suffer cosmic dilution.  If the extra
dimensions are themselves expanding or contracting, then branes
will obviously feel additional dilution/contraction.

Interestingly, according to (\ref{dilute}) branes which fill some
expanding directions such that $p_H \neq 0$, will feel less
dilution than completely wrapped branes ($p_H = 0$).  Thus
interactions between completely wrapped branes might be expected
to freeze out more quickly than between incompletely wrapped
branes ($p_H \neq 0$).

The kind of branes that can fill the spacetime will be determined by
the K-theory charges of the spacetime.  However, since D-brane charges
have yet to be classified for cosmological spaces, we will simply
assume that they are the same as in flat space, i.e., for Type IIA/B
theory all even/odd dimensional branes are possible, etc.  The kind of
cycles which branes can wrap and the number of such cycles are given
by the Betti numbers of the compact extra dimensions. The Betti
numbers largely dictate which brane configurations are possible.  For
example, a surface with $b_2=b_1=0$ possesses no one-dimensional or
two-dimensional cycles. Hence, $D2/D1$ branes will only be able to
exist as contractable two-dimensional/one-dimensional loops on the
surface.

A more common compactification like $K3\times T^2$ possesses
2-cycles from the K3 and $T^2$, two 1-cycles from the $T^2$,
3-cycles from combining a 1-cycle of the $T^2$ and a 2-cycle of
the K3, and 4-cycles by combining different 2-cycles of the K3 or
by combining a 2-cycle of the K3 with the 2-cycle of the torus.
Finally, 5-cycles also arise by combining the entire 4-cycle
of the K3 with a 1-cycle of the torus, and 6-cycles by combining
the entire  K3 with the 2-cycle of the $T^2$.  Thus
branes can wrap 1,2,3,4,5, or 6 dimensions.

Since branes and anti-branes will annihilate only if they are
parallel and share the same transverse space, we will assume that
all the branes are parallel and possess the same transverse space
$V_{\perp}$. We can then take the dimension of $n$ to be $[n] \sim
[L]^{-d_\perp}$.


We parameterize the annihilation rate $\Gamma$ as

\begin{equation}
\Gamma  = \frac{A_p}{m^{d_\perp-2}} \left(\frac{m}{T} \right )^{w}
\label{intrate}
\end{equation}

\n where $m$ is some mass scale and $A_p$ is roughly constant.

Although in this paper we will not worry about the functional
dependence of $A_p$, it will depend on the following.  It will
depend on the distribution of winding numbers of each cycle for
each type of brane/anti-brane.  This is because a brane and
anti-brane can annihilate only if their winding numbers,
$(n_1,...,n_w)$, around $w$ compact cycles are the same.
Collisions between branes can lead to inter-commutation of the
branes and a change in their winding numbers.  For example, a
brane wrapping say cycle $\gamma$ with winding number $w_{\gamma}
=1$ and a brane wrapping say cycle $\beta$ with winding number
$w_{\beta} = 1$ can collide and produce a single brane wrapping
both cycles $\gamma$ and $\beta$ with winding number
$(w_{\gamma},w_{\beta}) = (1,1)$ if the intersection number of
$\gamma$ and $\beta$ is nonzero. Thus the intersection numbers
tell us which winding number transitions are possible and $A_p$
will generally depend on them.

For example for the K3$\times T^2$ example, the intersection number
${}^\#(A_r \cdot B_s)$ of a $r$-dimensional cycle $A$ and a
$s$-dimensional cycle $B$ is one for two 1-cycles, $2b_2(K3)$ for a
2-cycle and 1-cycle.    Many intersection numbers are non-zero.  This means that branes wrapping
the various cycles of the compactification can collide to produce
branes with very different winding numbers. If some cycles are
isolated, then only branes wrapping those cycles can interact with
branes wrapping the same isolated cycles, reducing the volume
accessible to such branes.  This means that branes and anti-branes
wrapping identical isolated cycles interact more easily. However, if
the number of branes and anti-branes wrapping an isolated cycle are
different, then due simply to topology, complete annihilation of the
branes and anti-branes can not occur.

In general, collisions between branes wrapping compact cycles and
branes lying along non-compact directions will lead to branes
wrapping compact cycles and lying along non-compact
cycles~\cite{paper3}. Thus, spacetimes with both compact and
non-compact directions are unlikely to have branes lying along
only the compact or only along the non-compact directions.

$A_p$ will also be affected by the presence of any points/locii at
which the compact surfaces degenerate on the base manifold. At
such points the volume of the cycles decreases. Such points may
thus act as accumulation points for branes and anti-branes wishing
to minimize their energy.

For example, K3 possesses 24 singularities where a $T^2$ fiber
degenerates over a base ${\cal{B}}$ of the K3.  Thus, the two
dimensions of the branes wrapping the $T^2$ become very small at
the singularities.
Presumably at such singular points it is much easier for a brane
and anti-brane to annihilate, affecting $A_p$.

If some directions are fixed then  the number of expanding
dimensions satisfies $d_H < d$.  If some of the compact directions
are expanding then the expansion rate will be affected by the
wrapped branes.  For example, the negative pressure of a gas of
wrapped branes will increase the expansion rate.  If the dilaton
is allowed to freely evolve then the branes may actually inhibit
and halt the expansion~\cite{bv2}.  However, this halting will
only aid brane anti-brane annihilation if all of the directions
are compact, or if the branes are coincident in the non-compact
directions, or they fill the non-compact directions. In general,
expanding non-compact directions will still dilute brane densities
even if the compact directions stop expanding.

Thus, we take our spacetime to be a FRW universe. We fix the
dilaton. We ignore the expansionary effects of the brane density
as it increases the expansion rate and freezes out brane
interactions more quickly, making a FRW universe a conservative
choice.\footnote{Higher dimensional branes $\rho = -(p/d)
P_{ress}$.}  Then the Hubble rate, $H$, of such a radiation
dominated universe is:

\begin{equation}
H = -\frac{\dot{T}}{T} = \frac{T^{(d_H+1)/2}}{C m_p^{(d_H-1)/2}}.
\end{equation}

\n where $C^2$ is proportional to the effective number of spin
degrees of freedom.  Integrating (\ref{dilute}), we find
\cite{preskill}

\begin{eqnarray}
\frac{1}{f(T)} & = & \frac{1}{f(T_i) }+ \frac{AC}{w+p_H  +(1-d_H)/2} \left
(\frac{m_p}{m}\right)^{(d_H-1)/2} \nonumber\\
& {}& \times
\left(\left(\frac{m}{T}\right)^{ w + p_H +(1-d_H)/2} -
\left(\frac{m}{T_i}\right)^{ w +p_H +(1-d_H)/2}\right)
\label{numbranes}
\end{eqnarray}

\n where $f(T) \equiv n/T^{d_H-p_H}$ and is proportional to the comoving
density.  Here $T_i$ is the initial temperature at which annihilation
begins and $f(T_i)$ is proportional to the initial comoving density.
If $ w +p_H +(1-d_H)/2< 0$ then the annihilation is cut off for $T\ll T_i$.
If instead $ w + p_H +(1-d_H)/2>0$ as in (\ref{intrate}) then for $T\ll T_i$
the number of branes $f(T)$ becomes independent of the initial
temperature $T_i$ and approaches the limit

\begin{equation}
f(T\ll T_i) \approx
\frac{ w +p_H +(1-d_H)/2}{AC}\left(\frac{m}{m_p}\right)^{\frac{d_H-1}{2}}  \left(
\frac{T}{m} \right )^{ w +p_H +(1-d_H)/2}
\label{contannih}
\end{equation}

The annihilation rate will depend on brane anti-brane forces and brane
interactions with massless bulk fields like the graviton, dilaton,
etc.  If the bulk is thermalized, numerous closed string fields will
populate it.  As the branes move toward each other, bulk particles
will scatter off of them slowing them down.  To scatter a brane by
a large angle will require many bulk-particle scattering events, after
which the brane's velocity will be randomized apart from a drift
velocity in the direction of the attracting anti-brane.  Let $\lambda$
be the distance the brane can travel before it is scattered by a large
angle (mean free path).  Also let $r_c$ be the distance between a
stationary brane and anti-brane for which capture is almost sure.
Then, there are two important regimes: (1) the capture distance
exceeds the mean free path, $r_c > \lambda$; and (2) $r_c < \lambda$.

In the first case the branes will diffuse toward the anti-branes, and
once the branes reach within $r_c$ of the anti-branes they will
continue to be scattered such that all velocities transverse to the
direct path toward the anti-brane are randomized and effectively
zeroed away (impact parameter = 0).  The branes will thus collide
directly with the anti-branes.  If the drift velocity of a brane is
sufficiently slow, the brane will be captured by an anti-brane in a
bound state and eventually annihilation will occur.  Otherwise the brane
will simply pass through the anti-brane.\footnote{We thank Ian Kogan
and others for clarifications regarding solitons passing through one
another.}

Assuming collision occurs upon capture, the annihilation rate will be
the rate of branes diffusing toward the anti-branes.  The rate
per unit density $\Gamma$, will be

\begin{equation}
\Gamma = \frac{F}{n}
\label{annrate}
\end{equation}

\n where $F$ is the number of branes approaching the anti-branes per
unit time.  As in the diffusion of charged particles through a plasma,
this rate is

\begin{equation}
F = h^2 \left(\frac{\tau}{m_{Dp}}\right) n
\label{fluxvol}
\end{equation}

\n where $\tau$ is the mean free time and is the important quantity
determining the rate of brane collisions.

In the second case, once the brane is within $r_c$ of an anti-brane,
its transverse velocities will not necessarily be randomized and the
brane may streak past the anti-brane.  The only way for capture to
occur in this regime is for the brane to radiate away its kinetic
energy.  A moving brane drawn to an anti-brane will radiate via a
Larmor-type formula.  However, radiative capture is very weak and most
branes and anti-branes will survive as we show later.

At high temperature the bulk particles will be very energetic and the
mean capture length will exceed the mean free path (case 1).  However, as the
universe adiabatically expands, the temperature will drop until such
a time that brane-bulk interactions are frozen out at a
temperature $T_f$.  After $T=T_f$, the mean free path will exceed the
capture length and capture will occur only by radiative capture.
Hence, below $T_f$, brane anti-brane annihilation will become very
infrequent.

\section{Mean Free Path and Freezeout}

The capture distance $r_c$ can be obtained by the virial theorem,
KE$\sim$-PE.  Since ${\rm KE} \sim m v_{Dp}^2$, and PE$\sim
-h^2/r^{7-p}$, we find that

\begin{equation}
r_c = \left(\frac{h^2}{T}\right)^{1/(7-p)}
\label{capture}
\end{equation}

However, $D7$ branes are not asymptotically flat and possess a
logarithmically diverging inter-brane potential: PE$\sim -\ln
r$. Similarly, $D9$ branes have a quadratically diverging potential:
PE$\sim -r^2$. Thus, the capture lengths for $D7$ branes and $D9$ branes
are infinite.  In any case, the notion of capture length for $D9$
branes is not a meaningful since the branes fill the whole space.

We can obtain the mean free path, $\lambda$, for  a brane to be scattered by a large angle by bulk fields like gravitons, dilaton, etc., by writing
$\lambda = v_{Dp} \tau$, where $\tau$ is the mean time between
collisions in which the brane is scattered by a large angle and
$v_{Dp}$ is the brane velocity. If the branes move with thermal
velocities then $v_{Dp} = \sqrt{T/m_{Dp}}$ and $\lambda
=\sqrt{T/m_{Dp}} \tau$.

Now, in a time  $\tau$, the number of particles which are scattered is

\begin{equation}
N_s = v_a \tau \sigma n_a
\end{equation}
\noindent

\n where $\sigma$ is the cross section for a particle (graviton,
dilaton, etc,) to scatter off of a brane and $v_a$ is its average
velocity.  A particle can scatter two ways.  Suppose the brane
fills the $X^1,...,X^p$ directions.  Because the brane is an
extended object, a particle in the transverse directions may
directly strike the $Dp$ brane.  Since the brane has a width of
order $1/m_s$, unless the incident particle wavelength is smaller
than $1/m_s$ it will reflect off the brane (the transmission
coefficient is zero)~\cite{shellard}.  For a wavelength less than
$1/m_s$, the particle will pass through the brane.  Thus, for
temperatures below the string scale the cross section will be a
product of the geometrical size in the $X^1,...,X^p$ directions,
which is essentially the volume of the brane $V_p$, and the
cross-section to scatter off of it by streaking by it, $\sigma_s$.
Thus,

\beq
\sigma  = V_p \sigma_s ~~~~{\rm for} ~~ p < D-2
\eeq

\n Note, if $p=D-2$ (e.g. a $D8$ brane), $\sigma = V_p$.

Now, we want the time necessary for scattering by large angles.  A
particle can be scattered by a large angle when the amount of energy
$\Delta E$, imparted to it by thermal collisions is of order the mass
of the target object, in this case -- the brane mass, $m_{Dp}$. A gas of
thermal particles will impart an energy $\Delta E \sim T$ to the
target particle.  Thus the target particle will be scattered by a
large angle after approximately $m_{Dp}/T$ collisions with lighter
(massless) particles. Thus setting $N_s = m_{Dp}/T$ ~\cite{preskill}

\begin{equation}
\tau = \frac{1}{v_a \sigma n_a} \frac{m_{Dp}}{T}
\label{tautime}
\end{equation}

We have $v_a \sim 1$ and $n_a = a T^{D-1}$ since $D-1$ is the number
of spatial dimensions in which the light particles move. Thus

\begin{equation}
\lambda = \frac{1}{a \sigma} \frac{m_{Dp}^{1/2}}{T^{D-1/2}}
\label{lambda}
\end{equation}

\n Note that $\lambda$ grows with $m_{Dp}$ since the  heavier the
D-brane is, the harder it is to deflect by bulk scattering.

We can now write down an expression for the temperature at which the
mean free path equals the capture length, $T_f$.

Setting $r_c$ in eq \ref{capture} equal to $\lambda$ in
eq. \ref{lambda}, we find

\begin{equation}
T_f = \left (\frac{1}{h^2} \right )^x \left
( \frac{\sqrt{m_{Dp}}}{a \sigma(T_f)}\right )^y
\label{tFfinal}
\end{equation}

\n where

\begin{eqnarray}
x  =  \frac{1}{(D-1/2 )(7-p)-1} & \;\;\;\;\;\;&
y  =  \frac{7-p}{(D-1/2)(7-p)-1}
\label{exponent}
\end{eqnarray}

We parameterize the scattering cross section,
$\sigma_s$ as

\begin{equation}
\sigma_s = \frac{B}{m^{d_\perp-1}}\left (\frac{m}{T} \right )^{q}
\label{sigs}
\end{equation}

\n
where $m$ is some mass scale.  As section \ref{closedscat} shows for
closed string scattering off of branes, $q=2$ and $m=m_s$. That
calculation was done in the zero recoil limit.  For compact branes,
recoil will be more important and $q$ may deviate from its value $q=2$.

Using (\ref{sigs}), we find

\begin{equation}
T_f = \left (\frac{1}{h^2} \right )^{x/(1-qy)}
\left ( \frac{\sqrt{m_{Dp}}m^{d_\perp-q-1}}{aB V_p}\right )^{y/(1-qy)} \sim
\left (\frac{1}{V_p}\right)^{\frac{x+y/2}{1-qy}}
\label{tfinal}
\end{equation}

\n The exponent in (\ref{tfinal}) is

\beq
\frac{x+y/2}{1-qy} = \frac{1}{9.5 - \frac{1}{7-p} - q} \left
(\frac{1}{7-p} +\frac{1}{2} \right ) \approx
  \frac{1}{9.5 -  q} \left
(\frac{1}{7-p} +\frac{1}{2} \right )
\label{expone}
\eeq

The important feature of (\ref{expone}) is that the exponent is
strictly positive, ($p < 7$).  Using $q=2$ we find

\beq
T_f \sim \left (\frac{1}{V_p} \right )^{\frac{1}{7.5}(\frac{1}{7-p} +
\frac{1}{2}) }
\label{fout}
\eeq

\n To make the dependence of $T_f$ on $p$ more explicit, we set $V_p \sim
\ell^p$ and calculate $dT_f/dp$.  Here, $\ell$ is some length scale,
and the binary relation $\sim$ is used because the different
directions may have different sizes.

\beq
\frac{dT_f}{dp} \sim  - \frac{\ln \ell}{7.5}\left ( \frac{p}{(7-p)^2} +
\frac{1}{7-p} + \frac{1}{2}\right ) \left (\frac{1}{V_p} \right
)^{\frac{1}{7.5}(\frac{1}{7-p} +  \frac{1}{2}) } < 0
\eeq

\n Thus $dT_f/dp<0$.  This implies that the higher the dimension
of a brane (the smaller the co-dimension), the lower its freezeout
temperature is, since $V_p$ grows with $p$ and the exponent in
(\ref{fout}) also grows with $p$. This is particularly true if the
cycles the branes wrap, or non-compact dimensions the branes fill,
are much larger than the string scale.  Low dimension branes drop
out of thermal contact with the bulk fields first. Higher
dimensional branes continue to interact with bulk fields at lower
temperatures and are thus captured by the same dimensional
anti-branes at lower temperatures.  Intuitively, this is plausible
because the brane anti-brane potential grows with $p$, and hence
the capture distance grows with $p$.  Also, higher dimensional
branes can more easily interact with bulk fields because their
extended area is exponentially larger.  This is essentially a
selection rule which states that a universe populated with branes
will get rid of higher dimensional branes more easily.  We expect
this, since higher dimensional branes can meet much more easily in
the non-interacting case and presumably also in the interacting
case. This implies that co-dimension dominates over mass in the
annihilation process.  One might have expected larger mass objects
to drop out of equilibrium first.  However, because of the
extended size of the objects, this appears not to happen. For
branes wrapping small compact directions with radii $\sim
\sqrt{\alpha'}$, then $V_p \sim 1$, weakening the selection rule.
But, in most cases the radii will be at least tens of string
lengths. This is because brane formation is a stringy process
occurring around an energy  $\sim ~m_s$ (energy scale of tachyonic
phase transition). Because, of the energy in the momentum modes of
the strings and branes wrapping the compact dimensions, the cycles
will typically be expanding.  Thus by the time the branes have
formed and start to look for one another, the compact universe
will have substantially grown and the radii will be $\gg
\sqrt{\alpha'}$. Furthermore, if we assume that annihilation
occurs in the loitering phase of dilaton dynamics as in
\cite{bv3}, then radii$~\gg \sqrt{\alpha'}$.  This is because
loitering occurs after the compact universe has grown
substantially larger than its early string length size.

>From (\ref{fout}) we can also infer that for fixed $p$ that $T_f$
grows with the number of static dimensions, $p_s$, which a brane
fills. If we take all such static dimensions to be compact, then
$p_s$ is the number of dimensions a $Dp$ wraps. If we write $V_p
\sim \ell^p R^{p-p_s}$ where $R(t)$ is a cosmological expansion
factor, then

\begin{equation}
T_f \sim \left (\frac{R^{p_s}}{R^p\ell^p} \right )^{\frac{1}{7.5}(\frac{1}{7-p} +
\frac{1}{2}) }\ \ \Longrightarrow  \ \  \frac{\partial T_f}{\partial
p_s} \arrowvert_p > 0
\end{equation}

\n This is expected as (\ref{dilute}) shows that for fixed $p$
that $Dp$ branes are frozen out more quickly as $p_H$ decreases or
equivalently as $p_s = p-p_H$ grows.

Using equations \ref{annrate}, \ref{fluxvol}, \ref{tautime} we find
that

\begin{equation}
\Gamma = \frac{A_p}{m_s^{d_\perp-1}}
\left (\frac{m_s}{T}\right )^{D-2}
\label{rate}
\end{equation}

\n Since $ d_H -p_H \le D-1$ and $d_H \le D-1$, we find that the
condition for the number of $Dp$ branes to be independent of the
initial number: $ w +p_H +(1-d_H)/2>0$, is always satisfied since
$w = D-2$. Because, $T_f$ is essentially the lowest temperature at
which our annihilation mechanism works, we find that at late times
the number of $Dp$ branes is

\begin{equation}f_p(T_f) \approx
\frac{ D-2+p_H +(1-d_H)/2}{A_pC}\left(\frac{m_s}{m_p}\right)^{\frac{d_H-1}{2}}  \left(
\frac{T_f}{m_s} \right )^{ D-2 +p_H +(1-d_H)/2}
\label{finaldensity}
\end{equation}

Note that $T_f/m_s < 1$ and the part of the exponent in
(\ref{finaldensity}) which is $(T_f/m_s)^{p_H}$ decreases as
$p_H$ grows.

This means that the more expanding dimensions the branes fill, the
smaller their final quantities are.  If no compact directions
exist or the branes live only in the non-compact directions such
that $p=p_H$, then at freezeout the number of high dimensional
branes is lower than the number of low dimensional branes for
equal initial number of branes.  If compact directions exist and
the branes wrap $p_s$ directions, then since $p_H = p - p_s$,
fewer high dimensional branes still remain. Thus, in addition to the fact
that $T_f$ is lower for higher dimensional branes, this further
confirms that the number of higher dimensional branes is
exponentially suppressed compared to the number of lower
dimensional branes.

For a fixed $p$, since $p_H \propto -p_s$ as a $Dp$ brane wraps  more
directions its chance of being annihilated decreases.  This
is because according to (\ref{dilute}) the more compact directions
a gas of branes fills (the fewer compact directions it wraps) the
more diluted it becomes.  In the limit that the branes fill all of
the expanding directions such that $p_H = d_H$,
(\ref{finaldensity}) can be shown to take the form of $f_p(T_f)
\sim 1/\Gamma(T_f) t(T_f)$.

Because $f_p$ does not vanish remnants remain. This is troublesome
because once $T$ falls below $T_f$ annihilation can occur only by
radiative capture, an inefficient process. For example, a brane may be
attracted to an anti-brane.  However, to form a bound state, the
kinetic energy of both branes need to be radiated away.  In the case
of monopoles, it is known that almost all orbits leading to capture
are nearly parabolic (eccentricity = 1) at the point of closest
approach.  Most of the radiation is given off during closest approach
and the energy loss can change an initially hyperbolic orbit with
eccentricity slightly larger than unity to a bound state elliptic
orbit with eccentricity slighter smaller than unity.  Once captured in
this bound state, the brane then spirals in towards the anti-brane.
At stringy distances, tachyonic open strings appear leading to brane
anti-brane annihilation.

Branes and anti-branes possess Ramond-Ramond charges.  Thus, it is not
surprising that accelerating branes and anti-branes radiate and
possess a Larmor type formula\cite{abou}.  We write the total power
radiated by an accelerating brane/anti-brane as

\begin{equation}
P = \frac{1}{6 \pi}\gamma_{p,D} \Omega(D) h^2 \dot{v}^2
\end{equation}
\n $\gamma_{p,D}$ describes how the power radiated per unit solid
angle varies with $p$, and $\Omega(D)$ is the total solid angle in
$D$ spacetime dimensions; $\dot{v}$ is the acceleration which can
be found from the trajectory.

We will calculate the radiative capture rate for $D6$ branes
wrapped on six compact directions and moving in four non-compact
directions. Thus, from a four dimensional point of view, the $D6$
branes behave like point particles.  In this case, $D$ is
effectively equal to four and $p=6$. The general case of $Dp$
branes moving in $D-1$ dimensions is not much more interesting.
This is because lower dimensional branes have an inter-brane
potential which falls off faster and can less easily form bound
states. Also, higher dimensional branes like $D7$ branes are
expected to disappear quickly since the inter-brane potential of
$D7-\bar{D}7$ brane anti-brane pairs does not drop off with
distance, but grows logarithmically, $V \sim -\ln r$.

Because we are interested in only the gross properties of the
motion, we will approximate the trajectory for a given energy $E$
and angular momentum $l$ by the same trajectory as traced out by a
particle with the same energy and angular momentum in four
dimensions. We also approximate the orbit during closest approach
by  an ellipse with eccentricity near unity.

The radiation of kinetic energy leading to capture has been calculated
in the case of monopole anti-monopole annihilation before by
\cite{zelda,page,landau}, and the following calculations are virtually
identical.  The amount of energy radiated in a frequency interval
$(\omega, \omega + d\omega)$ is $dE_r = \omega N(\omega) d\omega$.
Here $N(\omega) d\omega$ is the number of quanta radiated with
frequency near $\omega$.  It can be parameterized as

\begin{equation}
N(\omega) d\omega = \frac{8h^6\gamma_{6,4}}{\pi l^2} (K_{1/3}^2(x) +
K_{2/3}^2(x)) x dx
\end{equation}

\n where $K_{1/3}, K_{2/3}$ are Bessel functions and $l$ is the
orbital angular momentum, and

\begin{equation}
x = \frac{l^3 \omega}{3 h^4 m} = \frac{v^2l^3\omega}{6h^4E_{k}}
\end{equation}

The expectation value of the total energy radiated is then

\begin{equation}
\langle E_r \rangle = \int_0^{\infty} N(\omega) d\omega = \frac{4
\gamma_{6,4}
\pi h^{10}}{v^2 l^5} E_k
\end{equation}

\n
where $E_k$ is the kinetic energy.  All the kinetic energy is radiated
away for angular momentum values less than a critical value, $l_c$

\begin{equation}
l_c  = h^2 \left(\frac{4 \pi \gamma_{6,4}   }{v^2}\right)^{1/5}
\end{equation}
\n
Capture will occur for $l<l_c$. This gives a classical cross
section of

\begin{equation}
\sigma = \pi \left (\frac{l_c}{\mu v} \right)^2
\end{equation}

\n
The thermally averaged interaction rate cross-section is therefore
(using $v \sim \sqrt{T/m_{D6}}$ )

\begin{equation}
\Gamma = \langle \sigma |v| \rangle
\approx  \frac{K}{m_{D6}^2} \left
(\frac{m_{D6}}{T}\right)^{\frac{9}{10}}
\label{xrssect}
\end{equation}

\n where

\begin{equation}
K =
(2^4 \pi^7 \gamma_{6,3}  )^{1/5} h^{4} \propto g_s^2
\end{equation}

\n Note that the mass term is not the string mass, $m_s$, but rather
the mass of the $D6$ brane, which is not exceedingly large because the $D6$
was taken to be wrapped on a compact surface.

Since in this case $p_H=0, d_H=3$, and $w= 9/10$, we have $w + p_H
+ (1-d_H)/2 <0$. We know from the discussion around
(\ref{numbranes}, \ref{contannih}) that the annihilation is
therefore cutoff and that the brane and anti-brane densities do
not appreciably change from their values at $T=T_f$. Thus, the
number of branes is not driven to zero and brane anti-brane
annihilation doesn't resolve the brane problem.  For the opposite
case where a $D3$ brane fills the expanding four-dimensional
spacetime and moves through the static six dimensional compact
space, the annihilation rate $\Gamma$ will be given by an
expression similar to (\ref{xrssect}) and the exponent $w$ is
expected to be positive. Then $w + p_H + (1-d_H)/2 = 14/10 +d_H/2
$ will be positive definite because $p_H=d_H$. This means that
because the $D3$ branes fill the expanding space their
annihilation is not cutoff and they continue to annihilate until
they completely disappear. This is despite the fact that
brane-bulk interactions are frozen out because the temperature
decreases as the universe expands.

\section{Catalogue of results}

Below we list a somewhat pedantic catalogue of results -- how our
results apply to different spacetimes. The spacetimes differ in
the number of compact, static and expanding directions.

\begin{enumerate}
\item {\em All directions non-compact and expanding:} Brane
interactions are frozen out and  branes are leftover. This
corresponds to the case where dynamical compactification of the
extra dimensions occurs after brane production.

\item {\em All directions non-compact and static:} Interactions do
not have to fight cosmic expansion. Thus all of the branes
disappear if the directions are static for a sufficiently long
period.  This may correspond to an initial state where all the
dimensions are initially non-compact, static and some directions
are later dynamically compactified.

\item {\em All directions non-compact, $d^{nc}_H$ are expanding,
$d^{nc}_s$ are static; branes fill only static directions:}  The
distance between branes and anti-branes  expands. Hence
interactions are frozen out and branes do not annihilate fast
enough. (This corresponds to the case of  $p \le d^{nc}_s$, or $p
\le 6$ if $d^{nc}_H=4$).

\item {\em All directions non-compact, $d^{nc}_H$ are expanding,
$d^{nc}_s$ are static; branes fill only the expanding directions:}
This corresponds to $p = d^{nc}_H$ or $p=3$ for $d^{nc}_H=4$. As
the branes are separated only in the non-expanding compact
directions, they do not feel the expansion and interact unhindered
by the expansion. Thus if non-expanding directions are static for
a sufficiently long period, virtually all of the $D(d^{nc}_H-1)$
branes  disappear.

\item {\em All directions non-compact, $d^{nc}_H$ are expanding,
$d^{nc}_s$ are static; branes fill $p_H$ expanding directions and
fill $p_s$ static directions:} As long as the branes move in some
expanding directions they do not annihilate efficiently.  This
only happens if $p_H <d^{nc}_H$.

\item {\em All directions compact and static:} There is no
expansion for interactions to fight. All the branes disappear if
the compact directions are static for a sufficiently long period.

\item {\em All directions compact and expanding:} Expansion will
eventually kill off interactions and brane anti-brane remnants
remain. However, if as in~\cite{bv2,bv3} the dilaton is allowed to
vary, then via some dilaton dynamics the branes may be able to
halt the expansion of the compact directions for a period
sufficiently long for them to annihilate.

\item {\em All directions compact, $d^c_H$  are expanding, $d^c_s$
are static; branes wrap only static directions:} The branes
effectively behave like point particles moving in a non-compact
spacetime.  They do not disappear.

\item {\em All directions compact, $d^c_H$ are expanding, $d^c_s$
are static; branes fill only expanding directions:} This is again
the case of expanding branes moving in a static space. The branes
disappear.

\item {\em All directions compact, $d^c_H$ are expanding, $d^c_s$
are static; branes wrap $p_H$ expanding directions and wrap $p_s$
static directions:} The compact directions which the branes wrap
will act to only dimensionally reduce the branes which move in the
expanding directions.  This then reduces to scenario 7.

\item {\em There are $d^c$ compact directions, $d^{nc}$
non-compact directions. The compact directions are static, and the
non-compact directions are expanding. } Branes with $p<d_{nc}$
feel the expansion and do not disappear.  For $p \ge  d_{nc}$
then: (a) if the branes completely fill the expanding directions
(e.g. non-compact $D3$ branes in a 4+6 split of spacetime) then
they  disappear; (2) if the branes only partially fill the
expanding directions and wrap $p_w$ directions such $p-p_w
<d_{nc}$ then the branes feel the expansion and create a remnant
problem; (3) in the unlikely case that all of the branes are
coincident in the non-compact directions the brane separation is
oblivious to the expansion and the branes  disappear.

\item {\em This is the most general case. There are $d^c$ compact
directions, $d^{nc}$ non-compact directions. $d^c_H$ compact and
$d^{nc}_H$ non-compact directions are expanding.  $d^c_s$ compact
and $d^{nc}_s$ non-compact directions are static.}  Branes  move
in the expanding non-compact directions if $p<d^{nc}_H$, and thus
do not disappear.  If $p \ge d^{nc}_H$ then branes  still do not
dissappear if they do not completely fill the non-compact space,
i,e., if $p-p_w < d^{nc}_H$.  If at the same time the branes wrap
some of the expanding directions, then either the wrapped branes
must halt the expansion of the expanding compact directions or the
brane remnants will remain. If they do do halt the such expansion
then the brane density  tends to zero.

\item{\em The case of some compact contracting directions}
Contracting directions do not shrink distances in the expanding
directions. Thus contraction is unlikely to aid brane
annihilation. However, this deserves further investigation.

\end{enumerate}

\section{Conclusions}\label{conclusions}

D-branes are attractive brane-world candidates.  However, a number
of fundamental questions must be answered before they become
persuasive alternatives to the standard big bang model.  Several
of these are: why does  there appear to be only one braneworld?
Why is that braneworld four-dimensional and not say a $D5$ brane?
Why are $D3$ branes favored over higher or lower dimensional
branes? If braneworlds were littered throughout spacetime, they
would typically over-close the universe. Hence, either other
braneworlds are very far away, diluted by something like
inflation, or forced to be very rare because of some dynamical
anomoly/tadpole cancellation mechanism, or their interactions with
matter in the bulk is highly suppressed, or there is only one
brane-world -- ours.

We showed that it is very difficult to dynamically produce an
universe which migrates from a universe filled with many branes to
a universe filled with only a few branes.  Brane anti-brane
interactions are cutoff by the expansion rate and annihilation
after that point becomes very inefficient.
Dynamical brane anti-brane annihilation in a non-compact universe
is in fact more unlikely because branes and anti-branes will
typically not be parallel.  Hence, brane anti-brane collisions
will typically lead not to annihilation but a complicated
configuration of intersecting branes.\footnote{On a compact
surface, branes tend to align themselves so as to minimize their
energies so that non-intersecting branes wrapping the same cycles
become parallel. Thus this may not be true.}

However, in a universe with $d^c$ extra dimensions which have
already been stabilised, we showed that branes  may disappear if
they fill all of expanding directions, $p \ge d^{nc}_H$.  For four
non-compact directions, this means that $Dp$ branes with $p \ge 3$
may disappear.  However, if the branes do not wrap all of the
expanding directions, they will remain.  For example, even $D7$
branes may remain if they are separated in only one non-compact
expanding direction. Universes in which
the expanding directions are compact may become brane-less only if
the wrapped branes conspire to halt the expansion. Otherwise, they
suffer from the same problems that universes with expanding
non-compact directions suffer from.

We found one interesting result: higher dimensional branes tend to
annihilate fastest, leaving lower dimensional branes behind. This
is intriguing because it implies that in a scheme where branes are
dynamically generated as in a tachyonic phase transition that the
branes which survive and cause cosmological problems are low
dimensional branes like $D3$ and $D1$ branes (in type IIB theory).
Higher dimensional branes disappear.  This at least begins to
answer the question of why the universe may be filled with $D3$
branes as opposed to higher dimensional branes.  This means that
only one additional mechanism -- inflation -- may be needed to
make the brane-world idea much more persuasive.  Such inflation
would dilute the $D3$ branes, making an observer on any $D3$ think
that there is only one braneworld. Also, it would dilute any line
($D1$ branes) or point ($D0$ branes) defects on the surfaces of
the $D3$ branes.  Such inflation would also dilute any small
dimensional branes in the bulk as well.
It would be interesting to see whether this same result holds once we
also include interactions between $Dp-Dp'$ branes, where $p\neq
p'$. In general terms this problem is very complex as there will not
only be forces between branes of different dimensionalities, but
branes oriented at angles, and intersecting configurations attracting
other intersecting configurations. We did not include these
interactions in the paper as they are irrelevent to the central
question asked in this paper: do branes and anti-branes annihilate
fast enough to make themselves small components of the energy density
of the universe?

\section*{Acknowledgements}

We thank Carsten Van de Bruck, Fernando Quevedo and Ian Kogan for
comments.  A.C.D thanks PPARC. M.M also thanks the Cambridge
Commonwealth Trust, Isaac Newton Trust, DAMTP, Hughes Hall and PPARC for
financial support.

\section{Appendix: Closed String Scattering off of $D$-branes}\label{closedscat}

Closed strings fields such as the dilaton, graviton,
anti-symmetric tensor field, and massive modes can interact with a
brane by exchanging closed strings.  One can think of this as
spontaneously created open strings on the brane absorbing and then
re-emitting closed strings and then subsequently annihilating
themselves.  This is the $t$ channel point of view, see figure
\ref{branebulk}. The dual $s$ channel process is: a closed string
strikes the brane, creates an open string excitation which then
propagates along the brane and is eventually annihilated when it
emits another closed string~\cite{klebanov, garousi}.

We are concerned most with the $t$ channel because, the $t$
channel scattering amplitude blows up for small scattering angle.
This divergence is also common to scattering by other solitons
like monopoles.

\EPSFIGURE[l]{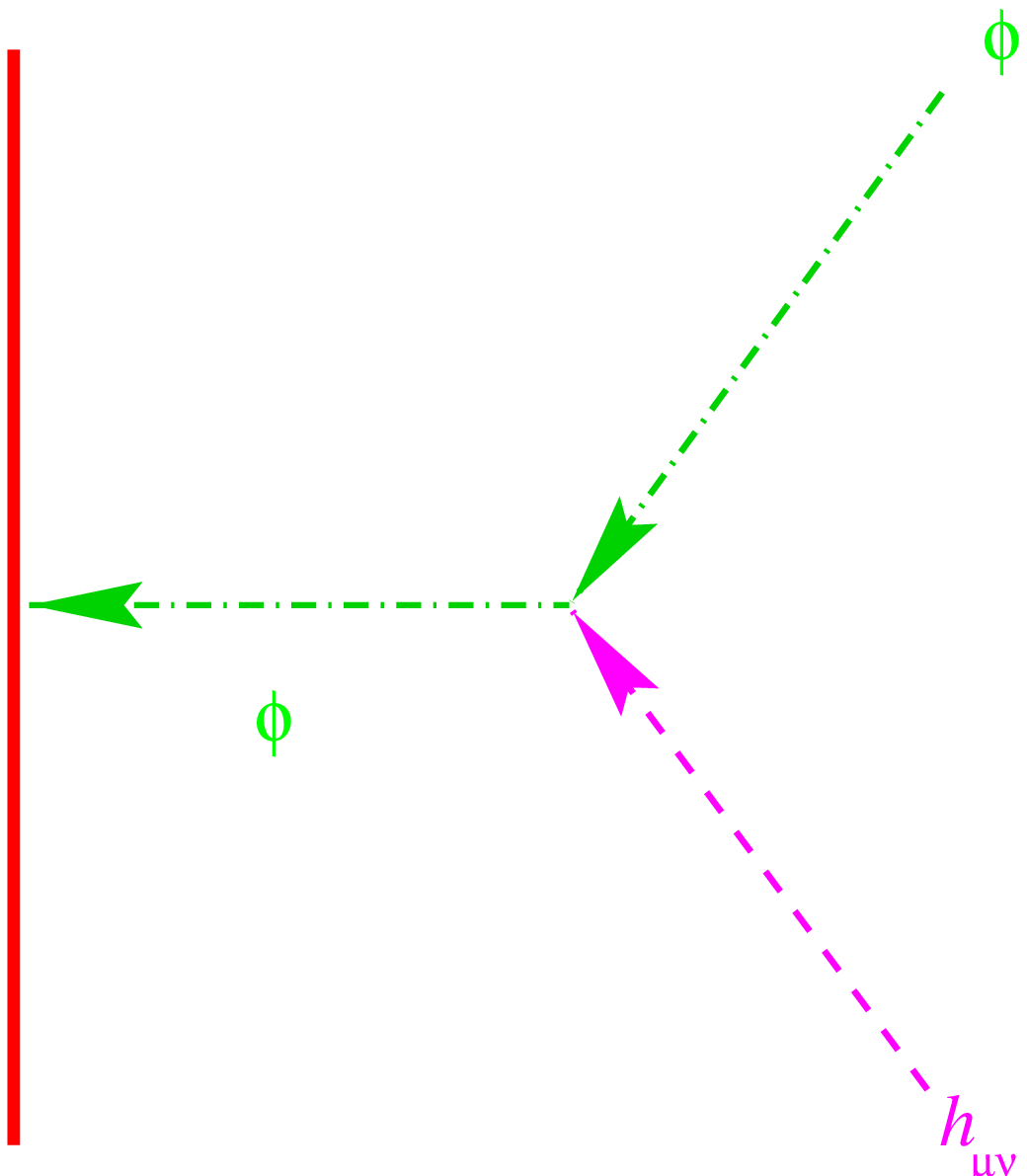,width=6cm}{$t$ channel scattering by a
brane. \label{branebulk}}

The form of the scattering amplitude can be guessed as follows.
Because of the exchange of a massless particle, there will be a
pole and $ {\cal{A}} \sim 1/t$ where $t$ is the exchanged momentum
squared. The numerator of the amplitude should possess some
momentum dependent factors because the interactions between the
bulk and the brane occur via quadratic derivative interactions.


Because the amplitude is stringy, there should also be a series of
Regge poles, allowing for not only massless closed string exchange, but
the exchange of massive closed strings as well.  Thus we expect that

\begin{equation}
{\cal{A}} =   \frac{a_1}{t} + \frac{a_2}{t+1} + \frac{a_3}{t+2} + \cdots
\label{tchn}
\end{equation}

\n where we have taken the poles, in appropriate units, to be
$n\in {1, 2 ,3....} = \IZ$. The $a_i$ factors are quadratic in the
initial and final momenta, $p_1,p_2$.  The amplitude for massless
exchange $a_1/t$, should be proportional the $g_s$, since for $g_s
\rightarrow 0$ the brane  should decouple from the bulk and no
scattering should occur.  The simplest way to achieve this is for
$a_1 \sim \kappa^2 \tau_p \cdot ({\rm function~of~}p_1,p_2),~$
since the amplitude should be proportional to Newton's constant
$\kappa^2 \sim g_s^2$.

Because momentum is conserved only parallel to the brane, (in the
limit of small recoil), the only invariants are

\begin{eqnarray}
s & = & 2 \alpha'(p^2_1)_{||} = 2 \alpha'(p^2_2)_{||} \nonumber\\
t & = & -\alpha'(p_1 +p_2)^2 = -2\alpha'p_1 \cdot p_2
\end{eqnarray}

More formally, in the limit of no recoil the tree level stringy amplitude can be
found by calculating the disk amplitude with insertions of two closed
string vertex operators in the interior of the disk.

\beq
{\cal{A}} = \int \frac{d^2z_1 d^2z_2}{V_{CKG}} \langle
V_1(z_1,\bar{z}_1) V_2(z_2, \bar{z}_2)\rangle
\eeq

\n $V_1$ and $V_2$ can be decomposed into two right moving
operators and two left moving operators. After fixing the
positions of three of them one finds the familiar Veneziano
scattering amplitude~\cite{klebanov, garousi}

\begin{equation}
{\cal{A}}  =   c_1\frac{\Gamma(t) \Gamma(s)}{\Gamma(1+s+t)} (s a_1 -t
a_2) \delta^{p+1}(p_1+p_2)
\label{venez}
\end{equation}

\n where $(s a_1 - t a_2)$ is a kinematic factor and is equal to $s {\rm
Tr}(\epsilon_1 \cdot \epsilon_2)$ if the initial and final state
polarisations, $\epsilon_1$ and $\epsilon_2$, are orthogonal to the
brane. The delta function insures that momentum is conserved on the
brane. When expanded in the $t$-channel ($t\ll 1, s\gg 1$), the
amplitude (\ref{venez}) takes the form of (\ref{tchn}).  The constant $c_1$ is

\begin{equation}
c_1 = \frac{1}{g_o^2 \alpha' } \cdot \left ( \frac{\kappa}{\alpha'}
\right )^2 \cdot (\alpha'^2) \sim g_s
\label{c1}
\end{equation}

The factor on the left is the normalization of the disk amplitude.
Note that because the open string coupling constant $g_o$ is
basically the Yang-Mills coupling constant, in dimensions other
than four it is dimensionful, $1/g_o^2 \sim \alpha'^2\tau_p$.  The
second factor in (\ref{c1}) comes from the insertion of two closed
string vertex operators. The third factor comes from evaluating
two sets of propagators like $\langle X(z_1) X(z_2) \rangle =
\alpha' \ln|z_1-z_2|$.\footnote{Note the conventions,
\begin{equation} g^2_o = \frac{1}{(2 \pi \alpha')^2 \tau_p};~~~~
\tau_p = \frac{1}{g_s (2\pi)^p\sqrt{\alpha'}^{p+1}};~~~~ \kappa^2
= \frac{(2\pi)^7}{2} g_s^2 \alpha'^4. \label{formulas}
\end{equation}}

The amplitude can also be derived from an effective gravity
picture~\cite{garousi}.  The NS-NS sector of the low energy action of
Type II theories is

\beq
I_{NS} = \int d^{10}x \sqrt{-g} \left[\frac{1}{2\kappa^2} R -
\frac{1}{2} (\nabla \phi)^2 - \frac{3}{2} H^2 e^{-\sqrt{2} \kappa
\phi} \right ]
\eeq

\n The source term  for a D-brane is

\beq
I_s = \int d^{10}x \sqrt{-g}[S^{\mu \nu}_B B_{\mu \nu} + S_{\phi} \phi
+ S^{\mu \nu}_h h_{\mu \nu}]
\eeq

To leading order the source functions will be delta function sources at
the positions of the branes.  For example, the Fourier transform of
the dilaton source is $\tilde{S}_{\phi}(k) = -\tau_p(p-3)/\sqrt{8}$.
If one expands the metric as $g_{\mu \nu} = \eta_{\mu \nu} + 2 \kappa
h_{\mu \nu}$, we can read off the  dilaton-dilaton-graviton vertex
$\tilde{V}_{h\phi\phi}$ as follows


\beq
2\kappa h_{\mu \nu} \partial^{\mu} \phi \partial^{\nu} \phi
\rightarrow 2\kappa \epsilon_{\mu \nu} p_1^{\mu} p_2^{\nu} \phi(p_1)
\phi(p_2)
\eeq

\n and find that the amplitude

\beq
{\cal{A}} = i\tilde{S}_{\phi}(t)\tilde{G}_{\phi}(t)\tilde{V}_{h\phi\phi}
\eeq

\n is the same as the massless part of (\ref{tchn}) and
(\ref{venez}) if we perform a field redefinition $h_{\mu \nu}
\rightarrow 2 \kappa h_{\mu \nu}$ and $ \phi \rightarrow \sqrt{2}
\kappa \phi$.

The high energy limit, $ s,t\gg 0$ of the amplitude (\ref{venez})
is exponentially damped because of the Regge poles (${\cal{A}}
\sim s \exp(-\alpha' s m(\phi)/2)$~\cite{gross}. However, it peaks
in the small momentum transfer limit $t\ll 1$, but large energy
limit $s\gg 1$.  In that case

\beq
{\cal{A}} = c_1 \frac{s}{t} = \frac{c_1}{2 \sin^2 \frac{\theta}{2}}
\eeq

\n where we have assumed that momentum of the incoming particles
to the brane such that $s=\alpha' E^2$. This is the massless term
in our heuristic derivation (\ref{tchn}). The above amplitude is
the zero recoil limit.  In this case, $p_1 = p_2$, because as in
usual Compton scattering for $m_{Dp} \rightarrow
\infty$~\cite{peskin}

\beq
\omega_2 =  \frac{\omega_1}{1+ \frac{\omega_1}{m_{Dp}} (1-\cos \theta
)}  \overbrace{\longrightarrow}^{m\rightarrow \infty} \omega_1
\eeq

Thus $t = p_1 \cdot p_2 = \alpha' E^2 (1-\cos \theta)$.  Because
of this and momentum conservation, the final state $|f \rangle $ will
be identical to the initial state $|i \rangle $.  The probability of
scattering is

\beq
{\cal{P}}(i \rightarrow i)= \frac{|\langle i | S |i \rangle
|^2}{\langle i | i \rangle
\langle i | i \rangle}  =   \frac{|{\cal{A}}|^2}{2 p_1 \cdot 2 p_1 }
\sim \frac{g_s^2}{E^2 \sin^2 \theta/2}
\eeq

\n where we have explicitly included the normalization, $\langle i | i
\rangle = \langle p_1 | p_1 \rangle = 2 E_{p_1}$.

The cross section will be proportional to ${\cal{P}}(i \rightarrow i)$
once we average over the angle $\theta$.  Thus

\beq
\sigma_s \sim \frac{g_s^2}{E^2} \frac{1}{2\pi} \int_{\theta_{min}}^{2\pi}
\frac{d\theta}{\sin^2 \theta/2} \sim  \frac{g_s^2}{E^2}
\label{xsect}
\eeq

\EPSFIGURE[l]{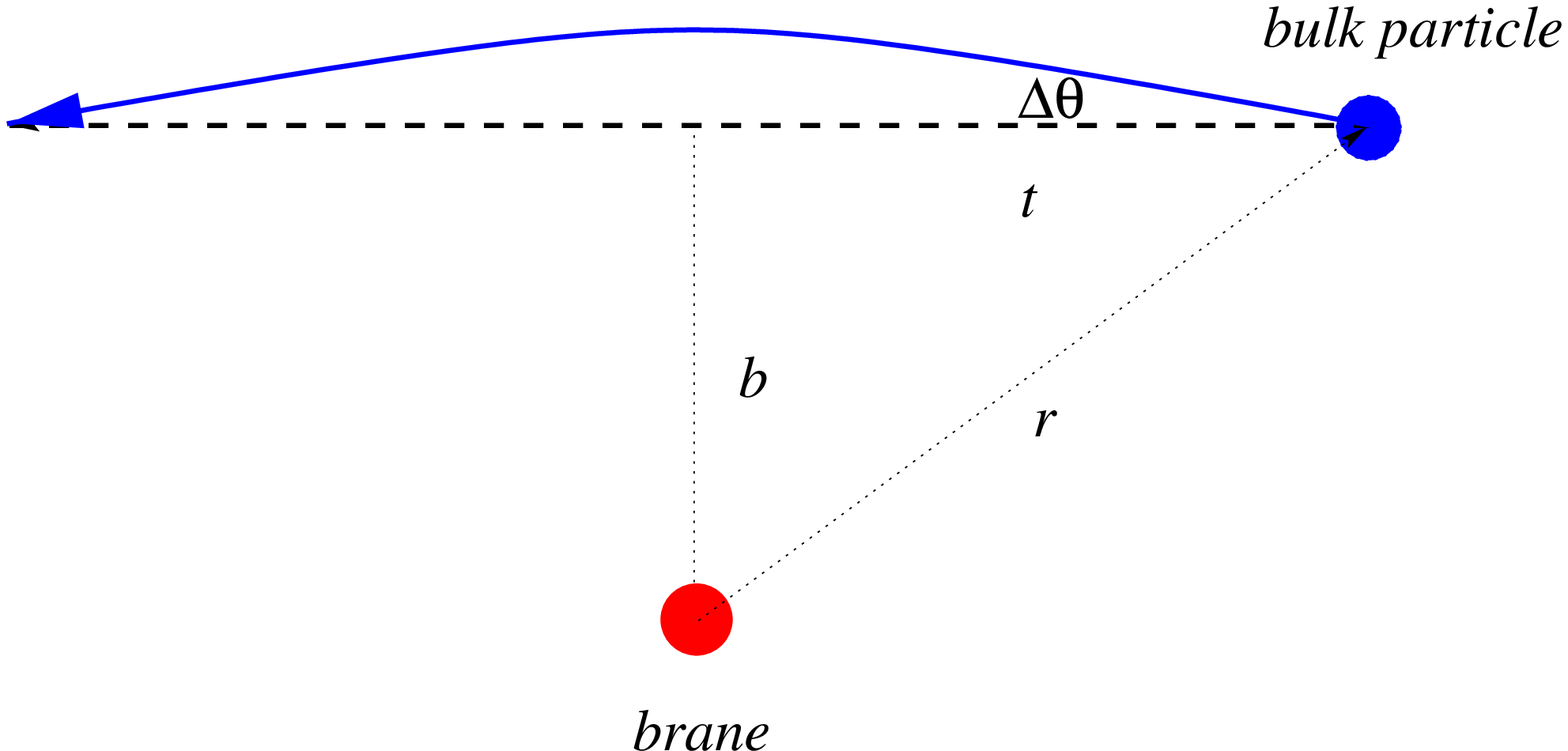,width=6cm}{Scattering of a bulk particle by
a brane.  Note, $t$ is time, and $r$ is the distance from the brane to
the bulk particle. \label{monopole}}

The dimensionful factors ignored in (\ref{xsect}) are simply factors
of $\alpha'$ and are not so important in determining the energy
dependence of $\sigma_s$.  The integral in (\ref{xsect}) diverges for
$\theta_{min} =0$.  Now a small deflection angle corresponds to a
large impact parameter~\cite{kolb2}. However, because the bulk fields
scatter off other bulk fields as well as the brane, the maximum impact
parameter of a particle which solely scatters off the brane is
bounded.  A bulk particle will feel a long range force from other bulk
particles.  The momentum change due to brane-bulk scattering will be
(see figure \ref{monopole})

\begin{eqnarray}
\Delta p & = & \int F dt \sim  \int \frac{dt}{r^{8-p}} =
\int_0^{\infty} \frac{dt}{(b^2 +t^2)^{(8-p)/2}} \nonumber\\
& = &
\frac{1}{b^{7-p}} \int_0^{\infty} \frac{d(t/b)}{((t/b)^2+1)^{(8-p)/2}}
\sim \frac{1}{b^{7-p}}
\end{eqnarray}

\n where $b$ is the impact parameter and $t$ is time. The angle of
deflection will be $\theta = \Delta p/p$.  If the mean free path of
the bulk fields is $\ell$, and $p\sim T$, then $\theta_{min} \sim
1/(T\ell^{7-p}).$ Hence, the forward scattering limit of (\ref{xsect})
is not singular.


\end{document}